\documentclass[sigconf, screen]{acmart}  %

\AtBeginDocument{%
  }

\acmConference[HotCarbon'24]{3rd Workshop on Sustainable Computer Systems}{July 9, 2024}{Santa Cruz, CA}
\acmDOI{}
\acmISBN{}
\setcopyright{rightsretained}

\usepackage{tikz}  %
\usepackage{todonotes}
\usepackage{xcolor}
\usepackage{url}
\usepackage{hyperref}
\usepackage{csquotes}
\usepackage{censor}
\usepackage[frozencache,cachedir=.]{minted}
 
\newcommand{\parag}[1]{\vspace{2mm}\noindent\textbf{#1.}\hspace{2mm}}

\newcommand*\circled[1]{\tikz[baseline=(char.base)]{\node[shape=circle,draw,inner sep=1pt] (char) {#1};}}

\begin{document}

\title{Vessim: A Testbed for Carbon-Aware Applications and Systems}

\author{Philipp Wiesner, Ilja Behnke, Paul Kilian, Marvin Steinke, Odej Kao}
\affiliation{%
  \institution{Technische Universität Berlin}
  \city{Berlin}
  \country{Germany}
}

\renewcommand{\shortauthors}{Wiesner et al.}

\begin{abstract}
To reduce the carbon footprint of computing and stabilize electricity grids, there is an increasing focus on approaches that align the power usage of IT infrastructure with the availability of clean energy. 
Unfortunately, research on energy-aware and carbon-aware applications, as well as the interfaces between computing and energy systems, remains complex due to the scarcity of available testing environments. 
To this day, almost all new approaches are evaluated on custom simulation testbeds, which leads to repeated development efforts and limited comparability of results.

In this paper, we present Vessim, a co-simulation environment for testing applications and computing systems that interact with their energy systems.
Our testbed connects domain-specific simulators for renewable power generation and energy storage, and enables users to implement interfaces to integrate real systems through software and hardware-in-the-loop simulation.
Vessim offers an easy-to-use interface, is extendable to new simulators, and provides direct access to historical datasets.
We aim to not only accelerate research in carbon-aware computing but also facilitate development and operation, as in continuous testing or digital twins.
Vessim is publicly available: \url{https://github.com/dos-group/vessim}.

\end{abstract}



\maketitle

\section{Introduction}

The rapid adoption of compute-intensive technologies like machine learning and big data processing has accelerated the growing demand for computing power in recent years~\cite{Wu_SustainableAIMLSys_2022, Payal_NatureCarbonImpactAI_2020}.
While the capacity of hyperscale datacenters doubled within only four years~\cite{Synergy_HyperscaleCapacityDoubles_2021}, also their global energy usage has increased by 10 -- 60\,\% between 2015 and 2021~\cite{IEA_DCEnergy_2022}.
This shows that improvements in hardware and software efficiency are still able to mitigate significant portions of growth in computing demand~\cite{Masanet_RecalibratingGlobalDCEnergyEstimates_2020}.
However, further efficiency improvements are becoming increasingly challenging, suggesting that future growth will likely translate into increasing energy demand more directly~\cite{IEA_DCEnergy_2022, Bashir_CaseForVirtualizingEnergySystem_2021, Masanet_RecalibratingGlobalDCEnergyEstimates_2020}.

\begin{figure}[t]
    \centering
    \vspace{2mm}
    \includegraphics[width=1\columnwidth]{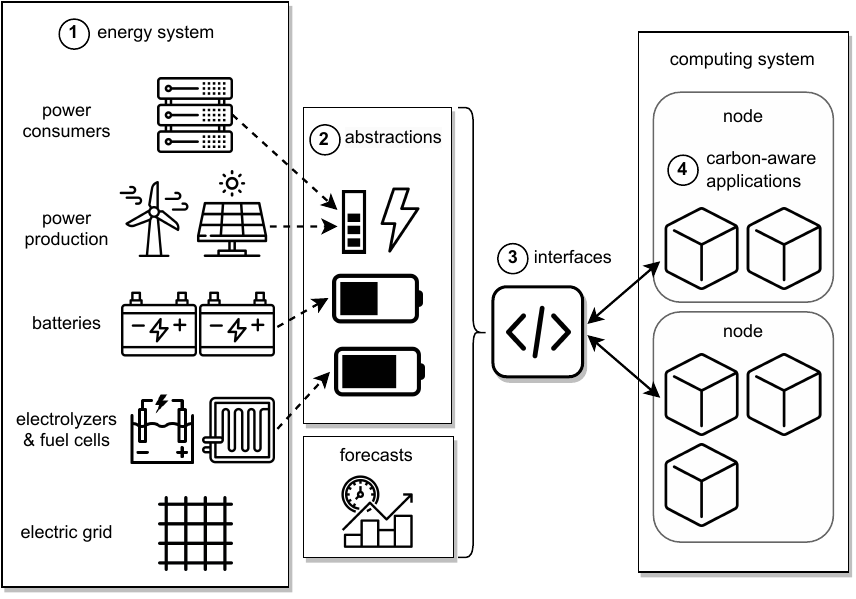}
    \Description{Different aspects regarding interactions between computing and energy systems.}
    \vspace{-5mm}
    \caption{Vessim enables research on various aspects regarding interactions between computing and energy systems.}
    \vspace{-3mm}
    \label{fig:1}
\end{figure}

To decrease the carbon footprint of computing despite increasing energy demand, a new paradigm called \emph{carbon-aware computing} has received much attention in academia~\cite{Radovanovic_Google_2022, Anderson_Treehouse_2022, Hanafy_CarbonScaler_2023, Wiesner_LetsWaitAwhile_2021} and industry~\cite{Google_CarbonAware, XBox_CarbonAware, VMWare_CarbonAware, Windows_CarbonAware}. 
Carbon-aware computing aims at aligning the power usage of compute infrastructure with the availability of variable renewable energy sources like solar and wind, for instance, by deferring or migrating flexible workloads. %
To perform carbon-aware decisions, we need up-to-date information on the datacenter's energy system -- which increasingly are microgrids comprising renewable power generation and energy storage.
However, today's energy systems are designed to hide underlying complexities~\cite{Bashir_CaseForVirtualizingEnergySystem_2021, Souza_Ecovisor_2023}, so relevant metrics are often not available to applications.
Examples include the real-time volume of on-site power generation, battery state of charge, or the current carbon intensity (gCO$_2$/kWh) of the public grid.
To enable researchers and developers to make their approaches carbon-aware, we need to improve the interplay between future energy and computing systems and create novel interfaces for information exchange.
For example, Souza~et~al. proposed \emph{ecovisors} to virtualize the energy system and provide software-defined access to applications~\cite{Souza_Ecovisor_2023}.

\emph{Nonetheless, research and development of applications that interact with energy systems, as well as the design of abstraction layers and interfaces, remains challenging due to the scarcity of available test environments.}
Carbon-aware approaches often involve globally distributed scenarios, where evaluations must be carried out over multiple days or months.
Only large industrial players can afford to test their approaches on real energy and computing infrastructures~\cite{Radovanovic_Google_2022} and due to the complexity and cost of constructing hardware testbeds, as well as their limited applicability for distributed settings, only a few approaches are evaluated on such~\cite{Souza_Ecovisor_2023, Liu_BatteryAging_2017}.
As a result, the majority of existing works simulate at least parts of the energy system~\cite{Hanafy_CarbonScaler_2023, Wiesner_LetsWaitAwhile_2021, Beldiceanu_EnergyProportionalCloudsRenewableEnergy_2017, Goiri_GreenSlot_2011, Goiri_GreenHadoop_2012, Goiri_ParasolAndGreenSwitch_2013, Zheng_MitigatingCarbonLoadMigration_2020, Liu_RenewableCoolingAwareWorkloadManagement_2012, Liu_GreeningGeographicalLoadBalancing_2011, Dodge_CarbonIntensityAICloudInstances_2022, 
Hanafy_GoingGreenLessGreen_2024,
Lin_CouplingDatacentersPowerGrids_2021, Kim_DataCentersHarnessStrandedPower_2017, Agarwal_VirtualBattery_2021, SEPIA_GreenITscheduling_2018, SEPIA_NegotiationGameITEnergyManagement_2020, wiesner2024fedzero}, for example, by replaying carbon intensity or solar production traces.
This leads to repeated software development efforts by researchers and little to no comparability between approaches.
Although first approaches use model-based analysis for planning carbon-aware datacenters~\cite{Acun_CarbonExplorer_2023}, there exist no solutions for actually testing their runtime behavior.

To effectively develop, operate, evaluate, and test car\-bon-aware ap\-proach\-es, we see the need for a modular and extensible testing environment. 
In this paper, we propose Vessim, a co-simulation testbed that combines domain-specific simulators for the different aspects of energy systems and enables the integration of real applications and hardware.
Our contributions are:

\begin{itemize}
    \item we propose a co-simulation design for connecting different simulators for power generation and energy storage to realistically represent microgrids for computing systems
    \item we explain how real computing systems can be connected to Vessim through software/hardware-in-the-loop simulation
    \item we demonstrate the usability of our testbed in an example scenario
\end{itemize}

Our goal is to facilitate and accelerate the development of carbon-aware applications and systems by enabling rapid prototyping and reproducible testing.

\section{Case for a Co-Simulation Testbed}

The testing of computer systems in simulated or emulated environments is commonly encountered where software depends on the behavior of physical systems, such as in the Internet of Things (IoT), edge computing, or cyber-physical systems~\cite{cintuglu2017, ValLedesma_SmartWaterTestBed_2021, VSimRTI_Mosaic_2011, MockFog_2021, Behnke_Hector_2019, Fogify_2020, Beilharz_Marvis_2021}. 
The modular nature of co-simulation allows the usage of established domain-specific simulators and has led to open standards like the Functional Mock-up Interface (FMI)~\cite{FMI} and various co-simulation frameworks~\cite{Mosaik_2019, CoSimulation_HELICS_2017, VSimRTI_Mosaic_2011}.

\parag{Research opportunities}
Using a co-simulation testbed enables research on various aspects related to the interplay of energy and computing systems, as shown in Figure~\ref{fig:1}:

\begin{itemize}
    \item[\textbf{\circled{1}}] \textbf{Energy system composition.}
    A co-simulation testbed facilitates rapid and cost-effective experimentation to determine optimal configurations for renewable power, batteries, or hydrogen storage~\cite{SEPIA_DATAZERO_2019} in datacenters. It enables parallel assessment of various compositions and exploration of otherwise expensive or inaccessible technologies.

    \item[\textbf{\circled{2}}] \textbf{Energy system abstractions.}
    Microgrids involve diverse producers, consumers, and storage types. While many complexities are irrelevant to carbon-aware applications, some are safety-critical. Configurable testbeds are valuable for experimenting with new abstractions, such as virtualizing the energy system~\cite{Bashir_CaseForVirtualizingEnergySystem_2021}.

    \item[\textbf{\circled{3}}] \textbf{Energy system interfaces.}
    Our testbed enables research on new interfaces covering
    (i)~the integration of new components and abstractions,
    (ii)~the handling of external data like weather forecasts,
    (iii)~the coordination of geo-distributed computing and energy systems, and
    (iv)~security concerns when exposing the energy system.

    \item[\textbf{\circled{4}}] \textbf{Carbon-aware applications.}
    Reproducible testbeds allow developers to prototype carbon-aware approaches quickly.
    In particular, we aim at providing access to current standards and best practices~\cite{SCI} and encourage the use of common datasets and baselines for evaluations.
\end{itemize}

\parag{Opportunities in development and operation}
Besides facilitating research, Vessim can support the development and quality assurance of carbon-aware applications and systems.
For example, it can be applied in continuous integration testing, or used to validate software roll-outs in a controlled environment.

Vessim can furthermore enhance the operation of carbon-aware datacenters by serving as a digital twin to predict future system states and help with decision-making.
This allows the integration of autonomous control strategies based on game theory~\cite{SEPIA_NegotiationGameITEnergyManagement_2020} or reinforcement learning~\cite{IEEETran_RenewableEnergyBigDataReinforcementLearning_2020, ICDCS_CloudResourceAllocationPowerManagementRL_2017}, which have been proposed in recent years.
Moreover, digital twins can be used for risk assessment in case of extreme events such as power outages.

\section{Vessim Overview}

\begin{figure*}
    \centering
    \begin{minipage}{0.7\linewidth}
        \includegraphics[width=\linewidth]{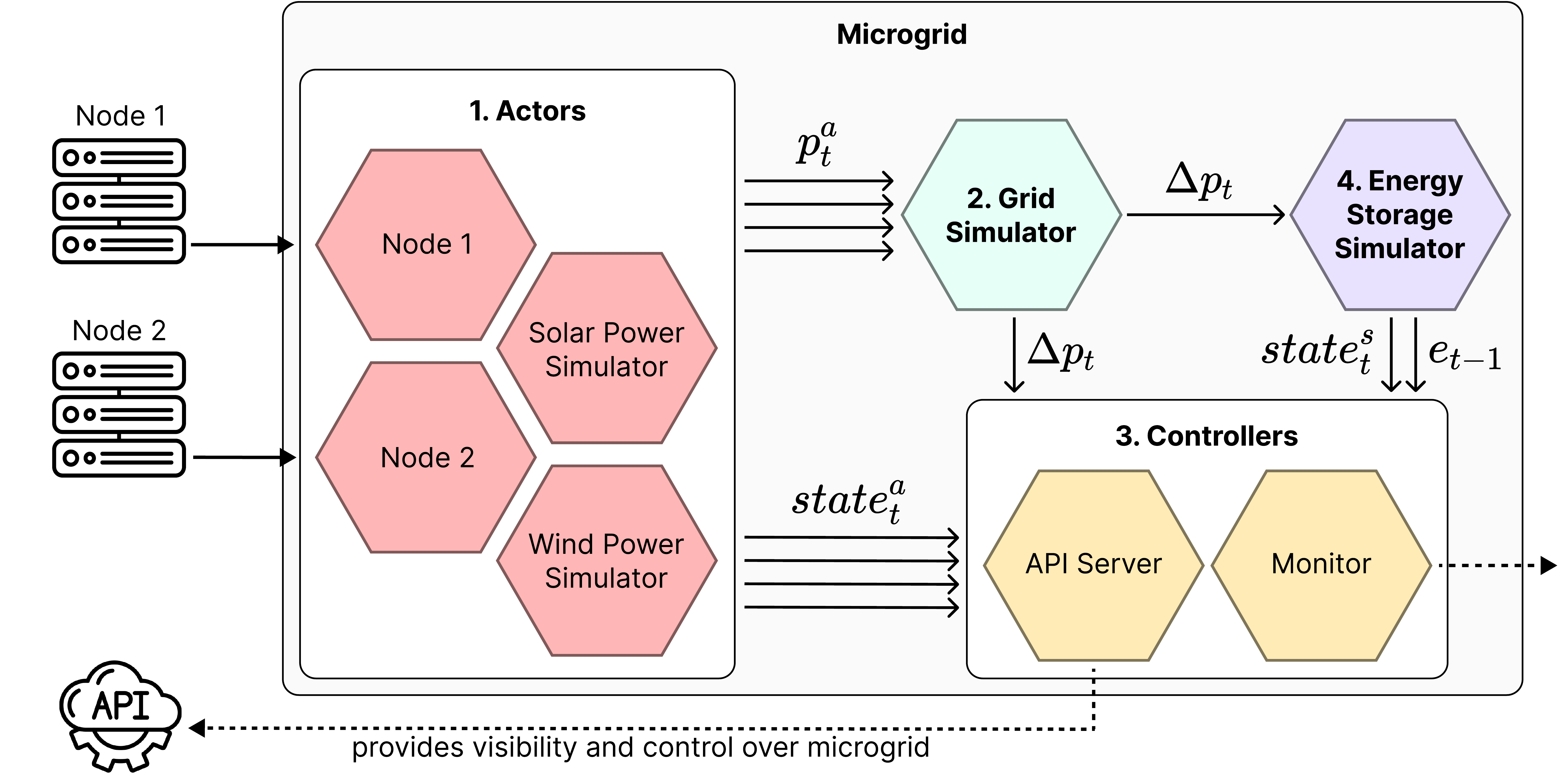}
    \end{minipage}\hfill
    \begin{minipage}{0.3\linewidth}
        \centering
        \small
        {\renewcommand{\arraystretch}{1.3}
        \begin{tabular}{|l|rrr|}
            \hline
            \boldmath{$t$} & 17:20 & 17:25 & 17:30 \\
            \hline
            \boldmath{$p^{\textbf{client}_1}$} & -480 W & -481 W & -485 W\\
            \boldmath{$p^{\textbf{client}_2}$} & -209 W & -170 W & -208 W\\
            \boldmath{$p^{\textbf{solar}}$} & 231 W & 241 W & 294 W\\
            \boldmath{$p^{\textbf{wind}}$} & 159 W & 151 W & 82 W \\
            \boldmath{$\Delta p$} & -299 W & -259 W & -317 W\\
            \textbf{battery soc} & 4.8 \% & 1.1 \% & 0\% \\
            \boldmath{$e$} & 0 Wh & 0 Wh & -17 Wh\\
            \hline
        \end{tabular}}
    \end{minipage}
    \caption{Vessim allows the simulation of multiple Microgrids in parallel where each consists of the depicted components. 
    	Hexagons represent a co-simulation subsystem managed by Mosaik~\cite{Mosaik_2019}.
    	\emph{Actors} represent consumers/producers; the \emph{grid} simulator determines the current excess/deficit in power; \emph{controllers} can be used for monitoring, decision-making, or providing APIs to computing systems; \emph{energy storage} simulators can charge/discharge batteries depending on user-defined policies.}
    \label{fig:overview}
\end{figure*} 

Vessim is based on the co-simulation framework Mosaik~\cite{Mosaik_2019}, which was originally developed for smart grid scenarios but has since its initial release in 2011 evolved into a comprehensive general-purpose co-simulator.
Mosaik offers APIs for connecting continuous and discrete-event simulators from any programming language as well as real systems through a TCP interface.
It furthermore enables users to specify which of these subsystems can interact during simulation runs.
At runtime, Mosaik coordinates the synchronization of subsystems and message passing.

Vessim provides a high-level Python API (§\ref{sec:example}) on top of Mosaik, which is designed to facilitate common use cases in the testing of carbon-aware systems.
In particular, users can define an arbitrary number of \emph{Microgrids} that each represents the energy system of a (distributed) computing system.
Vessim specifies a set of possible co-simulation components per microgrid (\emph{actors}, the \emph{grid}, \emph{controllers}, and \emph{energy storage}) which are all modeled by Mosaik subsystems and interact with each other as illustrated in Figure~\ref{fig:overview}.
The adaptable architecture of Mosaik facilitates easy extension of our framework.

Besides supporting the design of test scenarios through opinionated interfaces, Vessim offers a wide range of integrated simulators and ready-to-use datasets which are outlined in §\ref{sec:subsystems}.
Furthermore, Vessim enables software/hardware-in-the-loop simulation to integrate real applications and computing systems.
This capability is summarized in §\ref{sec:sil} and described in more detail in \cite{wiesner2023sil}.

The following sections explain the different types of co-simulation components that exist in Vessim.
The underlying Mosaik subsystems are executed in the following order: 1. \emph{actors}, 2. \emph{grid}, 3. \emph{controllers}, and 4. \emph{energy storage}.

\subsection{Actors}
Actors represent power producers and consumers within a microgrid.
At each simulation step $t$, each actor $a \in A$ relays its current power production ($p_t^a > 0$) or consumption ($p_t^a < 0$) to the grid simulator and, optionally, additional $state_t^a$ information to the controllers.
Examples of actors range from real servers equipped with power meters to entirely simulated systems, e.g. replaying previously recorded traces.
We provide an overview of already integrated subsystems in §\ref{sec:subsystems} and explain software and hardware-in-the-loop capabilities to integrate physical or virtual nodes in §\ref{sec:sil}.

\subsection{Grid}
Vessim's default grid simulator aggregates the power pro\-duc\-tion or con\-sump\-tion of all actors to determine the grid's current excess or deficit, i.e. $\Delta p_t = \sum_{a \in A} p_t^a$.

While this representation of power flows is accurate enough for most use cases, some scenarios require more precise modeling.
There are various tools for power system analysis like pandapower~\cite{Thurner_Pandapower_2018} or PyPSA~\cite{PyPSA} that are already integrated in Mosaik and can be used to periodically solve optimal power flows between actors.
Such modeling is useful when, for example, considering transmission loss over long distances or when testing approaches that optimize for curtailed energy~\cite{Zheng2022_MitigatingCurtailment,wiesner2024fedzero}, which is usually caused by congested grids.
Considering the expected impacts of growing datacenter power demands on the reliability of grids~\cite{Lin2024_GridPlanning}, power system analysis can also be valuable to investigate approaches that incorporate grid planning and management.

\subsection{Controllers}
Users can define a number of controllers to monitor the state of a microgrid and perform actions based on this information.
At every step $t$, each controller $c \in C$ receives all currently available information within the microgrid: $state_t^a$ $\forall a \in A$, the grid's $\Delta p_t$, the energy storage's $state_t^s$, and the energy surplus/deficit since the last step $e_{t-1}$ (see §\ref{sec:energy_storage_simulation}).
Additionally, users can define custom input signals to be used for decision-making.
For instance, the carbon intensity of energy from the public grid is a popular metric for scheduling carbon-aware applications.

Based on this information, controllers can persist the state of the microgrids over time, visualize this information in graphical user interfaces, or perform actions like throttling compute nodes or configuring the battery energy storage system.
A prominent use case of controllers is to provide API access to real computing systems to enable visibility and control over the microgrid~§\ref{sec:visibility_controll}.
An example scenario involving a custom controller can be found in §\ref{sec:example}.

\subsection{Energy Storage}
\label{sec:energy_storage_simulation}
While the above-mentioned components operate sequentially in the co-simulation, they all execute at the same step~$t$, effectively stepping "simultaneously."
In contrast, the energy storage simulator manages the charging and discharging of energy storage according to a user-defined policy, thereby modeling the progression of time between simulation steps.

The energy storage simulator receives $\Delta p_t$ along with the duration $d$ that passes between $t - 1$ and $t$.
A charge policy defines how $\Delta p_t$ is to be settled with the energy storage.
Any energy deficit/surplus that has not been settled between $t - 1$ and $t$ will be returned as $e_{t - 1}$, together with the new battery state $state_t^s$.
In other words, $e_{t - 1}$ represents the energy either drawn from ($e_{t - 1} < 0$) or fed into ($e_{t - 1} > 0$) the public grid.
If no storage is present, this simulator simply returns $e_{t - 1} = \Delta p_{t - 1} \cdot d$.

For instance, if $\Delta p_{t - 1} = -300 W$ and the previous step's duration is $d = 300s$, the default storage policy will try to discharge $300 W \cdot 300s = 25Wh$. 
In case only $10Wh$ are left in the storage, the simulator would return $e_{t - 1} = -15Wh$.
Users have the flexibility to configure the internal step size of the energy storage simulator and the level of realism in storage modeling, as detailed in §\ref{sec:storage}.

\section{Simulators and Datasets}\label{sec:subsystems}

This section describes which simulators and datasets are already integrated in Vessim, as well as possibilities for future extensions.

\subsection{Renewable Power Production}
\label{sec:renewable}

Vessim provides utilities for replaying historical weather and energy production data, as well as historical forecasts.
For example, we include two ready-to-use solar production datasets including forecasts provided by Solcast~\cite{Solcast}---one dataset covering ten cities in Germany with high geographic proximity and another dataset covering ten globally distributed cities.
If users require data from specific times and locations, they can utilize weather data services~\cite{OpenWeatherMap, Solcast, SolarAnywhere} that provide estimates of historical solar irradiance data (surface power density in W/m$^2$) worldwide.
This data can be multiplied by the panel area and efficiency factor (usually 15-20\,\%) and optionally factors like incidence angle, temperature, shading, soiling, or snow coverage~\cite{Milosavljevic_SolarSimulationReview_2022}.

For wind energy, we must consider factors like turbine type and efficiency, rotor diameter, blade pitch, and hub height.
For this, Vessim integrates the System Advisor Model (SAM) by the National Renewable Energy Laboratory of the US Department of Energy~\cite{NREL_Wind_2015} which can simulate single turbines as well as wind farms.

\parag{Next steps}
Besides solar and wind, the SAM provides simulators for various alternative renewable energy sources like concentrating solar power~\cite{NREL_ConcentratingSolarPower_2020}, marine energy~\cite{NREL_MarinePower_2013}, biomass combustion~\cite{NREL_Biomass_2011}, or geothermal power~\cite{NREL_Geothermal_2016}.
Furthermore, the hardware-in-the-loop interfaces deployed for feeding the power usage of actual servers to a simulation could also be used to inform the simulation about the power supply of real producers.

\subsection{Energy Storage}
\label{sec:storage}
Vessim supports modeling various types of energy storage with vastly different levels of realism. 
Alongside a simple battery model that does not consider any conversion losses, Vessim includes the C/L/C model for lithium-ion batteries~\cite{Kazhamiaka2019TractableLS}.
This model accounts for efficiency and imposes linear charging/discharging rate limits and has previously been used in related work~\cite{Acun_CarbonExplorer_2023}. 
For even more realistic battery modeling, we also provide an interface to the physics-based battery simulator PyBaMM~\cite{Sulzer_PyBaMM_2021}, which features an extensive range of battery models and simulation solvers.

\parag{Next steps}
We are currently assessing how different battery simulators affect experiment quality, runtime, and resource usage.
Additionally, Vessim could be extended by more diverse types of batteries.
For instance, the SAM~\cite{NREL_FuelCells} also includes simulators for lead-acid, lithium-ion, vanadium redox flow, and all iron flow batteries~\cite{NREL_Battery_2015} as well as models for battery degradation over time~\cite{Spotnitz_SimulationBatteryCapacityFade_2003, Smith_BatteryLifePrediction_2017}. Matlab/Simulink~\cite{Matlab_FuelCell} supports modeling of short-term energy storage options such as flywheels or supercapacitors~\cite{Matlab_Supercapacitor}, commonly used in uninterruptible power supplies.

Lastly, the idea to further reduce the reliance of datacenters on grid electricity through hydrogen storage, i.e. by providing them with fuel cells or electrolyzers, is prominent across industry~\cite{Microsoft_Hydrogen_2022} and research~\cite{NREL_HydrogenDataCenter_2019, SEPIA_DATAZERO_2019, Lazaar_HydrogenDataCenter_2021, Niaz_LeveragingOversupplyHydrogen_2022}.
Simulators for fuel cells can be found in the SAM and Matlab/Simulink.

\subsection{Historical Datasets and Live Data}\label{sec:datasets}

Next to the two solar datasets explained in §\ref{sec:renewable}, Vessim includes an exemplary dataset containing historical data and historical forecasts of the Marginal Operating Emissions Rate (MOER)~\cite{Watttime_LocationLocationLocation_2018}---a metric for describing the carbon intensity of grid energy---provided by WattTime~\cite{WattTime}.
Data providers like Solcast, WattTime, or Electricity Maps~\cite{ElectricityMaps} offer historical datasets for many regions around the globe which can directly be loaded within Vessim.
Furthermore, Vessim provides utilities for (i) returning perfect forecasts (i.e. returning the actual future data points), (ii) replaying historical forecasts, and (iii) including predictive models which generate forecasts during simulation.

\parag{Next steps}
Besides traces for weather, renewable power generation, or carbon intensity, there also exist various public datasets with cluster workload traces, such as from Alibaba~\cite{AlibabaClusterTraceProgram}, Azure~\cite{MicrosoftAzureTraces}, or Google~\cite{clusterdata:Wilkes2020a}.
When provided with appropriate power models that map infrastructure utilization to expected power demand, such traces can be replayed during co-simulations to model the power usage of large-scale computing systems.

Furthermore, we want to implement live data connectors to popular data providers or weather stations such as~\cite{HTW}.

\section{Software \& Hardware-in-the-Loop}
\label{sec:sil}
This section outlines how Vessim enables the integration of real computing systems into an energy system co-simulation, which is explained in more detail in~\cite{wiesner2023sil}.
The integration involves two steps: First, reflecting the computing system's power demand in the simulation. Second, enabling the computing system to exert visibility and control over the energy system.

It is important to note that \emph{as soon as real software or hardware is part of the simulation, we have to process the co-simulation in real-time} which is already supported by Mosaik~\cite{Mosaik_2019}.
However, for real-time simulations, we must ensure that all computations within a simulated time step take less time than the elapsed real time.
Complex subsystems that are prone to exceed this time limit should therefore be equipped with sufficiently powerful---or even multiple---machines. 
On the other hand, if \emph{all} components of a Vessim experiment are simulated, the speed of the co-simulation is mainly determined by the complexity of its subsystems and the performance of the host system.
Users can assess months or years of simulated time within minutes of real-time, which can be useful to, for example, investigate the impact of battery aging.

\subsection{Computing System Power Demand}

Vessim enables the integration of physical or emulated hardware through special \emph{actors} which periodically query and relay the real or emulated power demand.

\parag{Physical nodes}
For physical nodes, there exist different ways of estimating the power usage of hardware, like Intel's \emph{Running Average Power Limit (RAPL)} technology.
However, if precise power demand measurement is required, it is recommended to resort to external energy meters as most devices cannot accurately measure their power consumption using only software.
Our current prototype queries power demand directly from a metered power distribution unit, but we are already working on actors that retrieve relevant information from monitoring systems like Prometheus~\cite{Prometheus}.

\parag{Virtual nodes}
For virtual nodes, like VMs or containers, users need to define one or more metrics that are considered significant indicators of the node's power consumption (like CPU usage).
Using power models, these metrics can then be translated into the virtual node's power usage.
Power models can be based on mathematical models~\cite{ComputingServerPowerModelingSurvey_2020} or benchmarks such as SPECpower~\cite{SPECpower}.
As with physical nodes, this data can usually be queried from existing monitoring systems.

\parag{Overhead}
Power demand for cooling, uninterruptible power supply, or lighting is traditionally reflected in a datacenters power usage effectiveness (PUE)~\cite{ISO_IEC_30134-2_2016} %
which can be incorporated into the corresponding actor.
For more precise modeling of overhead, users can resort to advanced power models~\cite{Radovanovic_PowerModelng_2021} or, even better, measure the power demand of their real computing system beforehand under different conditions.

\subsection{Enabling Visibility and Control}\label{sec:visibility_controll}

As illustrated in Figure~\ref{fig:overview}, Vessim comes with a \emph{controller} that serves a REST API, similar to the one proposed in ecovisor~\cite{Souza_Ecovisor_2023} which provides applications with visibility (e.g. the current carbon intensity and solar production) and control (e.g. setting battery charge rates or throttling the power usage of applications) over the simulation.
The exposed information and API design are fully configurable by users, see points \circled{2} and \circled{3} in Figure~\ref{fig:1}.

When applications interface with a real-time simulation, there is a risk that the simulation will become overloaded in certain situations and fall behind schedule (see the introductory comment in §\ref{sec:sil}).
For example, a battery simulation may exceed its maximum processing time per simulation step if a large number of applications send requests to update their virtual battery configuration.
To avoid this, Vessim's API server comprises a buffer that collects and aggregates \texttt{SET} requests to forward them to the simulation at pre-defined intervals.
Likewise, the server includes a cache containing the relevant state of the energy system, ensuring that \texttt{GET} requests do not disrupt the simulation.
See~\cite{wiesner2023sil} for more details.

\section{Example Scenario}
\label{sec:example}

We demonstrate the current capabilities and usability of Vessim through an example scenario set in an edge computing environment that involves software-in-the-loop simulation.
Listing~\ref{listing} shows how this scenario is implemented using Vessim's Python API.

\begin{listing}
\begin{minted}[linenos,
               breaklines,
               numbersep=5pt,
               fontsize=\small,  %
               frame=lines,
               framesep=2mm]{python}
import vessim as vs

class CustomController(vs.Controller):
    ...

environment = vs.Environment()
environment.add_microgrid(
  actors=[
    vs.Actor("jetson", signal=vs.HttpPowerMeter( "https://127.0.0.2:8080/power"),
    vs.Actor("solar", signal=vs.HistoricalSignal.load( "solcast2022_global", column="San Francisco", params={"scale": 50}))
  ],
  storage=vs.SimpleBattery(capacity=250, initial_soc=0.6, min_soc=0.5),
  controllers=[
    vs.Monitor(outfile="result.csv"),
    vs.CustomController(moar=vs.HistoricalSignal.load( "watttime2022"), column="CAISO_NORTH")
  ],
  step_size=60, # 1 min
)
environment.run(until=24 * 3600)  # 24 h
\end{minted}
\vspace{-2mm}
\caption{Scenario definition with Vessim’s Python API. The scenario connects a solar and battery simulation with a physical, power-metered device (NVIDIA Jetson TX2).}
\label{listing}
\end{listing}

\parag{Actors}
After creating an \mintinline{python}{vs.Environment}, we add a single microgrid to the simulation.
This microgrid contains two \emph{actors}:
The first actor contains a \mintinline{python}{vs.HttpPowerMeter} that periodically queries the power usage of a single physical node (NVIDIA Jetson TX2) which is connected to a metered power distribution unit.
The Jetson is running a stress test at full load during the entire experiment.

The second actor represents a small solar panel with 50\,W maximum output. Production is modeled using a \mintinline{python}{vs.HistoricalSignal} that represents the "San Francisco" location in the "solcast2022\_glo\-bal" dataset that is included in Vessim.

\parag{Grid and energy storage}
Our example uses Vessim's default grid simulator that simply aggregates the current production/con\-sum\-ption from all consumers.
The microgrid is equipped with a \mintinline{python}{vs.SimpleBattery} with 250\,Wh capacity that has an initial state of charge (SoC) of 60\,\%. 
The configured charge policy allows a minimum SoC of 50\,\%.

\begin{figure}[h]
    \centering
    \vspace{2mm}
    \includegraphics[width=1\columnwidth]{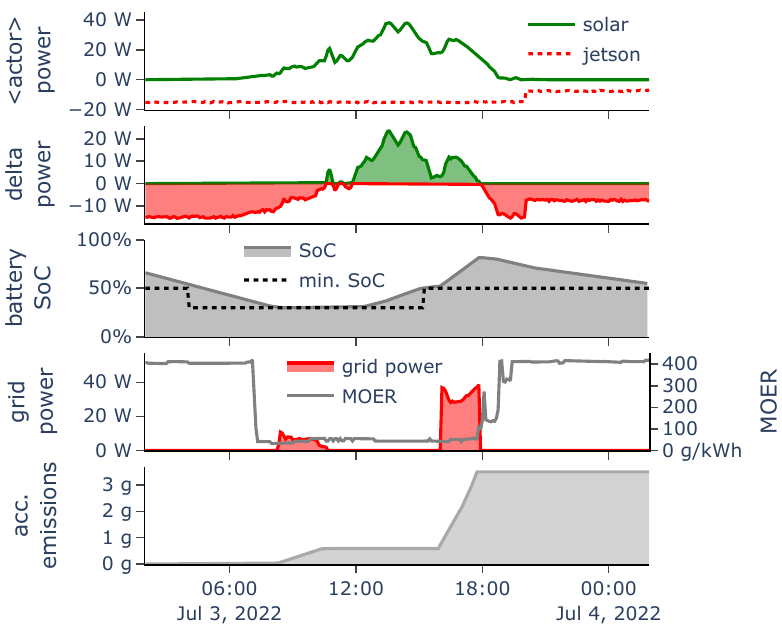}
    \vspace{-3mm}    
    \caption{Results of the example scenario. 1. $p^a_t$ of both actors, 2. $\Delta p_t$ computed by the grid simulator, 3. the battery's SoC (part of $state_t^s$), 4. energy deficit ($-e_t$) \& MOER, and 5. the accumulated carbon emissions ($-e_t \cdot$ MOER).}
    \label{fig:results}
\end{figure}

\parag{Controllers}
We instantiate two controllers: First, we configure a \mintinline{python}{vs.Monitor} to record all simulation state in a CSV file. The results are visualized in Figure~\ref{fig:results}.

Second, we implement a custom controller which has access to another \mintinline{python}{vs.HistoricalSignal}, representing the MOER in the \mintinline{python}{CAISO_NORTH} region where San Francisco is located.
Due to space limitations, we do not provide the controller's code in this paper, but describe its actions and their effect on the simulation:

\begin{enumerate}
    \item Once the battery reaches its minimum SoC, the controller queries renewable power production forecasts (historical forecasts are part of the "solcast2022\_global" dataset that the actor replays). Due to promising predictions, it temporarily lowers the minimum SoC to 30\,\% to avoid consuming carbon-intensive grid energy (MOER above 400\,gCO2/kWh).
    \item Once 30\,\% SoC is reached, the microgrid draws energy from the now comparably clean grid (MOER below 60\,gCO2/kWh).
    As evening approaches, the controller adjusts the charge policy to maintain a constant charging rate of 40\,W to refill the battery.
    This requires drawing more energy from the grid, as displayed in the 4th and 5th line of Figure~\ref{fig:results}.
    \item At 6pm, the microgrid transitions back to island mode, meaning that is does not draw energy from the public grid anymore. 
    To ensure overnight operation, at 8pm, the controller activates power-saving mode on the Jetson, reducing its power demand from {\raise.17ex\hbox{$\scriptstyle\sim$}}15 W to {\raise.17ex\hbox{$\scriptstyle\sim$}}7.5 W.
\end{enumerate}

This scenario demonstrates Vessim's user-friendly programming interface and many of its currently implemented features.
Our proposed testbed enables interactions between real computing systems and a fully simulated energy system.

\parag{Further examples}
Our paper on software-in-the-loop simulation for carbon-aware applications~\cite{wiesner2023sil} presents a similar, yet more detailed scenario based on Vessim, which includes physical and virtual nodes. 
FedZero~\cite{wiesner2024fedzero} does not incorporate software-in-the-loop simulation but leverages Vessim to simulate multiple distributed energy systems in parallel, faster than real-time.

\section{Conclusion}

While research in sustainable computing is gaining momentum, many studies overlook real-world complexities due to the lack of testing environments.
To fill this gap, we presented Vessim, a versatile co-simulation testbed for energy-aware and carbon-aware applications and systems, which connects energy domain-specific simulators with real software and hardware.

\emph{We invite researchers across the computing and energy disciplines to participate in the development of a common co-simulation testbed.}
Our goal is to accelerate progress in carbon-aware computing and research on datacenters' impact on power grids by continuously integrating the latest ideas from both fields.

In addition to the many opportunities for future work outlined in this paper, our primary focus will be on validating and calibrating Vessim simulations against real systems.
Another valuable direction would be to extend Vessim to cover further aspects of energy systems beyond electricity, such as water and thermal management.

\begin{acks}
We sincerely thank our student assistant Amanda Malkowski for her help on implementing the current prototype of Vessim.
This research was supported by the German Ministry for~Education and Research (BMBF) as Software Campus (grant 01IS17050) and \mbox{BIFOLD} (grant 01IS18025A).
\end{acks}

\bibliographystyle{ACM-Reference-Format}
\bibliography{bib-with-dois}

\end{document}